\documentclass[aps,prb,twocolumn,groupedaddress]{revtex4-1}

\usepackage{graphicx}
\usepackage{dcolumn}
\usepackage{bm}
\usepackage{amsmath}
\usepackage{amssymb}
\usepackage{latexsym}
\usepackage{epsfig}
\usepackage{amsbsy}
\usepackage{array}
\usepackage{amssymb}
\usepackage{setspace}
\usepackage{bm}
\usepackage{subfigure}
\usepackage{caption}

\def\sint{\ifmmode{- \!\!\!\!\!\! \int}
    \else{\hbox{$- \!\!\!\! \int \ $}}\fi}

\renewcommand{\raggedright}{\leftskip=0pt \rightskip=0pt plus 0cm}

\begin{document}


\title{Andreev spectroscopy of the triplet superconductivity state in Bi/Ni bilayer system}

\author{Xin Shang}
\author{Haiwen Liu}
\author{Ke Xia}
\email[Ke Xia: ]{kexia@bnu.edu.cn}
\affiliation{Department of Physics, Beijing Normal University, Beijing 100875, China}


\begin{abstract}
We calculate the Andreev spectroscopy between a ferromagnetic lead and a Bi/Ni bilayer system. The bilayer system is described by  Anderson-Brinkman-Morel(ABM) state and mixing ABM and $S$-wave state.
In both the ABM state and the mixed ABM state and $S$-wave state, the Andreev conductance is consistent with that obtained in the point contact experiment[Zhao,et al, arXiv:1810.10403].
Moreover, the conductance peak near the zero energy is induced by the surface state of the ABM phase.
Our work may provides helpful clarification for understanding of recent experiments.
\end{abstract}



\maketitle

\section{Introduction}  

Triplet $p$-wave superconductors
have received much interest\cite{Tinkham} which provide new insights into topological superfluidity\cite{Vollhardt1991The,doi:10.7566/JPSJ.85.022001},
superconductivity, and new spintronics applications\cite{PhysRevLett.111.226801,RevModPhys.80.1083}.
Especially, topological $p$-wave superconductors\cite{RevModPhys.82.3045,PhysRevLett.102.187001,PhysRevB.81.094504,PhysRevB.79.214526,1367-2630-12-6-065010} promise quantum computing applications such as  Majorana fermions which locate at the edges and the vortex cores of superconductors\cite{PhysRevB.61.10267,PhysRevLett.86.268,Volovik2009,Volovik20091,PhysRevLett.105.186401,PhysRevB.73.220502}.
Topological $p$-wave superfluid of $^3He$ have been reported\cite{PhysRevLett.103.155301} and superconductivity in $Sr_2RuO_4$  has been suggested\cite{PhysRevLett.107.077003,KASHIWAYA201425}.
Another peculiar feature of topological materials is the  gapless  surface states\cite{PhysRevLett.49.405,PhysRevB.25.2185,doi:10.1143/JPSJ.81.011013,KASHIWAYA201425}.
Experimentally, the surface state can be detected by the Andreev spectroscopy\cite{PhysRevB.51.1350,Kashiwaya_2000,KASHIWAYA201425}.

Recent point contact experiments have observed triplet superconductivity in epitaxial Bi/Ni bilayers\cite{moodera1990superconducting,kumar2011physical,siva2015spontaneous,Gong_2015,liu2018superconductivity,zhao2018triplet,PhysRevB.99.064504}.
Triplet p-wave superconductivity was inferred from the zero-bias peak of the Andreev conductance  between the epitaxial Bi/Ni bilayer and the ferromagnetic metal.
Furthermore, a quantitative analysis of the Andreev  conductance revealed a triplet $p$-wave Anderson-Brinkman-Morel (ABM) state\cite{PhysRevLett.5.136,PhysRevLett.30.1108}, with two Weyl nodes.
In contrast, the recent time-domain THz spectroscopy experiment\cite{chauhan2019nodeless} have reported a nodeless bulk superconductivity in the epitaxial Bi/Ni bilayer.
In addition, the inversion symmetry of the Bi/Ni bilayer is broken, suggesting the mixing of different pairings such as $S$-wave and $p$-wave\cite{sato2017topological,bauer2012non}.
The Bi/Ni bilayer system naturally raises two questions:
(1)Does the broken inversion symmetry admit any superconducting paring other than the ABM state at the interface?
(2) Given the importance of the behavior of surface states in topological superconductors, how do those surface states and bulk ABM states contribute to the transport properties?

\begin{figure}
\scalebox{0.35}[0.35]{\includegraphics[35,29][590,694]{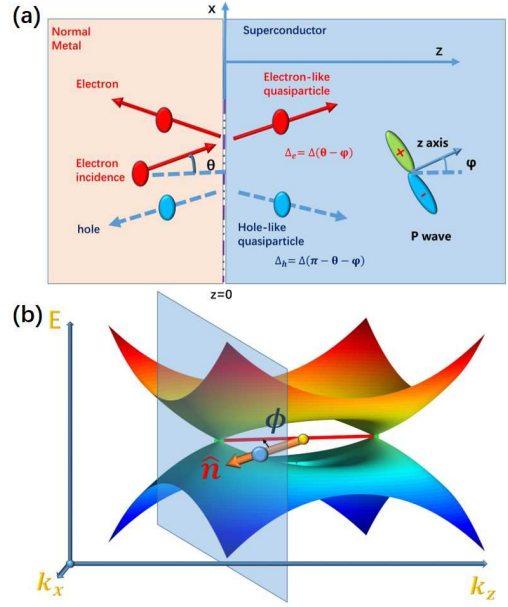}}
 \caption{\raggedright  Schematic diagram of (a) the transport system and (b) the band structure of the ABM state.
 $\phi$ represents the angle between the z axis of the $p$-wave and the normal to the interface.
The transmitted hole-like and electron-like quasiparticles have  effective pairing potentials of $\Delta_e=\Delta_p\sin(\theta-\phi)$ and $\Delta_h=\Delta_p\sin(\pi-\theta-\phi)$ respectively.
  \label{fig1}}
\end{figure}

We build a model that calculates the Andreev spectroscopy and local density of states of  superconducting materials i.e. the ABM state and the ABM state mixture with $S$-wave pairing.
We first  calculate the conductance of the Andreev reflection in pure ABM state using the Blonder-Tinkham-Klapwijk (BTK) method\cite{blonder1982transition,bauer2012non}, and its local density of states by the surface Green's function method\cite{PhysRevB.95.235143,PhysRevB.55.5266,0305-4608-15-4-009}.
Second, we calculate the Andreev conductance of the ABM state mixture with  $S$-wave superconductivity.
The Andreev conductance of both the pure ABM state and the mixed state with a small $S$-wave component were consistent with the results of point contact experiments\cite{zhao2018triplet}.
However, in mixed states with a  large $S$-wave component, the conductance deviated from the point contact results\cite{zhao2018triplet}. After computing the local density of states of those state, we find nodes in the pure ABM state and the mixed state with small $S$-wave component, but not in the mixed state with large $S$-wave component. The local density of states of mixed state with large $S$-wave component  is consistent  with the time-domain THz spectroscopy experiment\cite{chauhan2019nodeless}.
We also revealed that the conductance peak near the zero energy is contributed by the surface state.

The remainder of this paper is organized as follows.
Section~\ref{model}, introduces our model for calculating the conductance.
Section~\ref{conductance-a}, and ~\ref{conductance-p+s},  calculate the conductances and the surface statestates of the ABM state and an unconventional superconductivity state(an ABM state mixed with a $S$-wave state), respectively. The paper concludes with a brief summary.
\\

\section{Model }\label{model}
Consider a normal metal-superconductor (N-S) junction located at z = 0( where z $>$ 0 represents the superconductor, and z$<$0 represents normal metal) as shown in Fig.~\ref{fig1}(a). Here, we only consider the case of incident on the x-z plane, because the pairing function is isotropic on the x-y plane for the ABM state.

The effective Hamiltonian in the Nambu representation is given by\cite{bauer2012non}:
\begin{equation}
H_S=\left(\begin{array}{cc}
\hat{H}_{0}(k) & \hat{\Delta}(k)\\
-\hat{\Delta}^{*}(-k) & -\hat{H}_{0}^{*}(-k)
\end{array}\right),
\end{equation}
where $\hat{H}_{0}(k)=\zeta_{k}$, $\zeta_{k}=\frac{\hbar^{2}}{2m}k^{2}-\mu$,  $\hat{\Delta}(k)=i\Delta\sigma_{y}$ for singlet pairing and $\hat{\Delta}(k)=[\hat{d}(k)\cdot\vec{\sigma}]i\sigma_{y}$ for triplet pairing\cite{bauer2012non}.

We first consider the superconducting order parameter $\Delta_p$ of of a p-wave superconductor in the pure ABM  state\cite{PhysRevLett.5.136,PhysRevLett.30.1108}:
\begin{equation}
\hat{\Delta}(k)=
\left[
 \begin{matrix}
   -\Delta_p \sin\theta_k & 0\\
   0 & \Delta_p \sin\theta_k\\
  \end{matrix}
  \right].
  \label{abm}
\end{equation}
where $\sin\theta_k\equiv\frac{k_{Fz}}{k_F}$ with $k_F$ being the Fermi momentum and  $k_{Fz}$ being the $z$ component of $k_F$.
According to  BTK theory\cite{blonder1982transition,bauer2012non}, the wavefunction of a superconductor $\Psi_{s}$ is given by:
\begin{equation}
e^{ik_{x}x}[c_{1}\psi_{1}e^{iq_{1z}^{+}z}+c_{2}\psi_{2}e^{-iq_{2z}^{+}z}+c_{3}\psi_{3}e^{iq_{1z}^{-}z}+c_{4}\psi_{4}e^{-iq_{2z}^{-}z}],
\end{equation}
where $q_{1(2)z}^{\pm}=\sqrt{q^{2}_{1(2),\pm}-k^2_{Fx}}$,  $q_{1(2),\pm}^2
\approx k_F$.
Here, $\psi_{1(2)}=[u_{e(h)}^+,0,-v_{e(h)}^+,0]^T$ denotes the electron(hole) like state for spin index $\uparrow$, and $\psi_{3(4)}=[0,u_{e(h)}^-,0,v_{e(h)}^-]^T$ denotes the electron(hole) like state for spin index $\downarrow$, $u_{e}^{\pm}=\sqrt{\frac{1}{2}(1+\frac{\varepsilon_{e}^\pm}{(|E|)})}$,
$u_{h}^{\pm}=\sqrt{\frac{1}{2}(1+\frac{\varepsilon_{h}^\pm}{(|E|)})}$, $v_{e}^{\pm}=\alpha_{e}^\pm\sqrt{\frac{1}{2}(1-\frac{\varepsilon_{e}^\pm}{(|E|)})}$,
$v_{h}^{\pm}=\alpha_{h}^\pm\sqrt{\frac{1}{2}(1-\frac{\varepsilon_{h}^\pm}{(|E|)})}$, with $\alpha_{e(h)}^+=\alpha_{e(h)}^-=sign(sin\theta_{e(h)})$. Here, $\varepsilon_{e(h)}^+=\varepsilon_{e(h)}^-=\sqrt{(|E|)^2-(\Delta_{e(h)})^2}$, with $\Delta_{e(h)}=\Delta_p\sin\theta_{e(h)}$, $\theta_{e}=\theta_k-\phi$ and $\theta_{h}=\pi-\theta_k-\phi$ denote the effective pair potentials of electron-like and hole-like quasiparticles, respectively\cite{PhysRevLett.74.3451,bauer2012non,zhao2018triplet}, where $\theta_k$ depicts the electron incident angle and $\phi$ represents the angle between the x axis of the $p$-wave and the normal  to the interface  (similar to the  angle $\alpha$ between  the x axis of the d wave the interface normal in a d-wave superconductor\cite{PhysRevLett.74.3451}), as shown in Fig .~\ref{fig1}(b).

Second, we consider a mixed $S$-wave pairing and ABM state, whose superconducting order parameter has the following form:
$\hat{\Delta}(k)=-\Delta_p \sin\theta_k\sigma_z+\Delta_si\sigma_y$\cite{bauer2012non,PhysRevLett.92.097001}.
Then, the superconducting order parameters split into two independent order parameters $\Delta_{e(h)}^+ = \Delta_p\sin\theta_{e(h)} + \Delta_s$ and $\Delta_{e(h)}^- = \Delta_p\sin\theta_{e(h)} - \Delta_s$ respectively\cite{PhysRevB.76.012501}.
In this case, the wave function changes to
$\psi_{1}=[u_{e}^{+},u_{e}^{+},-v_{e}^{+},v_{e}^{+}]^T$ , $\psi_{2}=[u_{h}^{+},u_{h}^{+},-v_{h}^{+},v_{h}^{+}]^T$,
$\psi_{3}=[u_{e}^{-},-u_{e}^{-},v_{e}^{-},v_{e}^{-}]^T$ and $\psi_{4}=[u_{h}^{-},-u_{h}^{-},v_{h}^{-},v_{h}^{-}]^T$.
Here, $u_{e(h)}^\pm$ and $v_{e(h)}^\pm$ have the same form as the former case,
with $\varepsilon^{\pm}_{e(h)}=\sqrt{(|E|)^2-(\Delta_{e(h)}^\pm)^2}$,  $\alpha_{e(h)}^\pm=sign(\Delta_{e(h)}^\pm)$.

The wavefunction in the lead region  is derived from the
 Hamiltonian:
$\hat{H}_N(k)=\zeta_{k}-\mu+\vec{M}\cdot\vec{V}$;
where $\zeta_{k}$, and $\mu$ denote the kinetic energy and the chemical potential respectively.
The plane wave at the normal metal side can be expressed by a four-component wavefunction in the Nambu representation :
\begin{equation}
\Psi_{N}=e^{(ik_{F}x)}\left(\begin{array}{cc}
   e^{ik_{Fz}}+b_{\uparrow,\uparrow}e^{-ik_{Fz}}\\
   b_{\uparrow,\downarrow}e^{-ik_{Fz}-\gamma}\\
   a_{\uparrow,\uparrow}e^{ik_{Fz}}\\
   a_{\uparrow,\downarrow}e^{ik_{Fz}+\gamma}
   \end{array}\right).
\label{psin}
\end{equation}
The first row $e^{ik_{Fz}}+b_{\uparrow,\uparrow}e^{-ik_{Fz}}$ of Fq.\ref{psin} describes an electron with a spin
up incident plane wave and a normal reflection wave.
The second row $b_{\uparrow,\downarrow}e^{-ik_{Fz}-\gamma}$ describes an electron with a spin down wave.
The third row $a_{\uparrow,\uparrow}e^{ik_{Fz}}$ and the fourth row $a_{\uparrow,\downarrow}e^{ik_{Fz}+\gamma}$
are hole descriptors with a spin-up and a spin-down Andreev reflection wave, respectively.
The $\gamma=0$ in normal metal(NM) lead, and $\gamma=\infty$ in ferromagnetic metal(FM) lead\cite{PhysRevLett.109.146602} describe the evanescent wave.
Note that we only consider the incidence of spin-up electrons. Fully polarized ferromagnetic lead, contain only spin-up electrons whereas in nonmagnetic lead, the spin-up and spin-down electrons are identical, so it is sufficient to consider spin-up electrons only.

Next, we study the transport properties of the N/S junction.
We assume that the N/S interface located at z=0 along the x axis has an infinitely narrow insulating barrier described by the delta function $U=U\delta(z)$\cite{yokoyama2005intrinsically,bauer2012non,zhao2018triplet}.
Solving the following boundary conditions\cite{yokoyama2005intrinsically,bauer2012non,zhao2018triplet}£º
\begin{equation}
\begin{split}
& \varPsi_{S}(0)=\Psi_{N}(0)\\
& \frac{\partial \varPsi_{S}(z)}{\partial z}|_{z=0}-\frac{\partial \varPsi_{N}(z)}{\partial z}|_{z=0}=U\varPsi_{S}(0),
\end{split}
\end{equation}
we obtain $a_{\uparrow,\uparrow(\downarrow)}$ and $b_{\uparrow,\uparrow(\downarrow)}$.
The normalized conductance with a bias voltage is\cite{zhao2018triplet}:
\begin{equation}
\sigma(eV)=\frac{\intop_{-\pi/2}^{\pi/2}g^{T}(eV)cos\theta d\theta}{\intop_{-\pi/2}^{\pi/2}g^{T}(\infty)cos\theta d\theta}
\label{conductance}
\end{equation}
where $g^{T}(eV)=\intop_{0}^{1}g(|eV+\frac{1}{\beta}ln\frac{1-f}{f}|)$, and  $g(E)=[1+\frac{1}{2}\sum_{\rho=\uparrow,\downarrow}(|a_{\uparrow,\rho}(\theta,E)|^{2}-|b_{\uparrow,\rho}(\theta,E)|^{2})]$, $\beta=\frac{1}{k_{B}T}$, the parameter $\Gamma$  represents the energy broadening\cite{PhysRevLett.109.146602}.

\section{The conductance and surface state  of ABM state}\label{conductance-a}
\begin{figure}
\centering
\scalebox{0.42}[0.42]{\includegraphics[17,14][581,671]{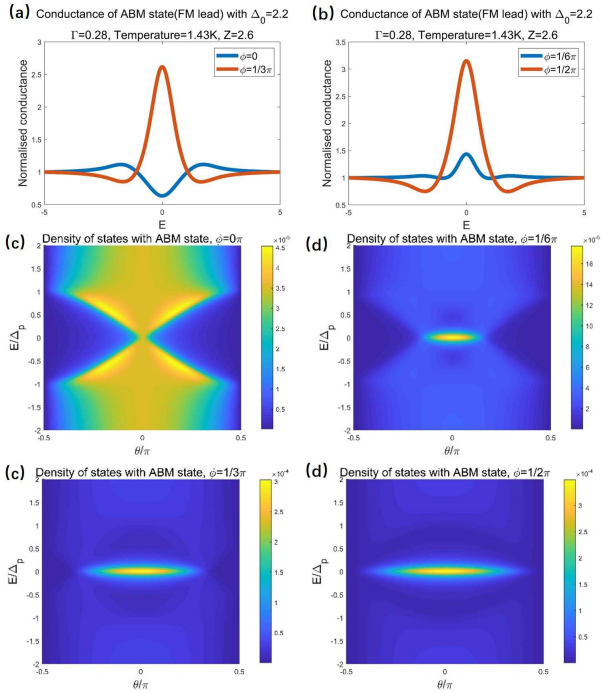}}
 \caption{\raggedright Conductance and the density of states for different incidence planes. Conductance of (a)$\phi=0^o$ and $60^o$, (b) $\phi=30^o$ and $90^o$; densities of states for(c)$\phi=0^o$, (d)$\phi=30^o$, (e)$\phi=60^o$,and (f) $\phi=90^o$. Here, $\theta$ is the incident angle, $\phi$ is represents the angle between the x axis of the $p$-wave and the normal to the interface.
As $\phi$ increases, the conductance near zero energy increases, and the surface state near the zero energy  becomes increasingly obvious.
The interfacial parameter was obtained by data fitting to the experimental data of the c face of FM lead.
  \label{fig2}}
\end{figure}
We first calculated the normalized Andreev conductance between ferromagnetic/non-magnetic lead and ABM state at different incident planes, denoted by $\phi$.
The interface parameters were obtained by fitting the conductance to the experimental results on the c plane. In subsequent analysis, we mainly set  $\phi$ is equal to $0^o$  and $60^o$ because  the results of $60^o$ approximated the experimental results of the c plane.
Detail information about experimental fitting procedure is detailed  in the Appendix.

As shown in Fig ~\ref{fig2}(a) and (b), the conductance near the zero energy changed from a valley to a peak as $\phi$ increase.
At $\phi$ is zero, the conductivity near the zero energy was valley shaped.
Increasing the $\phi$, gradually increases the conductance at zero energy, and the conductance peak.

Comparing the density of states localized on the surface with the energy band of the ABM state,  we observe that the conductivity at the zero energy is contributed by  the projection of the surface state between the two Weyl points on the incident plane at the zero energy.
First, as the local density of states on the surface was consistent with the conductivity spectrum, the conductance could be attributed to the strength of the density of states.
At $\phi=0^o$[Fig~\ref{fig2}(c)], the density of states exhibited a funnel-like shape, forming a valley of conductance.
As $\phi$ increased [Fig~\ref{fig2}(d)-(f)], the density of states became increasingly concentrated around the zero energy, leading to a more pronounced peak in the conductance spectrum.
However, as shown in Fig ~\ref{fig1}(b), the band structure of the ABM state was similar to that of  Weyl semimetals, with only two Weyl points at zero energy.
Previous work has reported a Fermi arc between the two Weyl points\cite{Su2015TOPOLOGICAL,Jia2016Weyl}.
Comparing the density of states with the band structure, we inferred that the zero energy state is the projection of the Fermi arc on the incident plane.

Conductance spectroscopy of the Andreev reflection between ferromagnetic/non-magnetic lead and the ABM state at different incident planes was consistent with the density of states.  Both the density of states and the conductance spectrum are given in the appendix.

\begin{figure*}
\centering
\scalebox{0.8}[0.8]{\includegraphics[22,20][592,239]{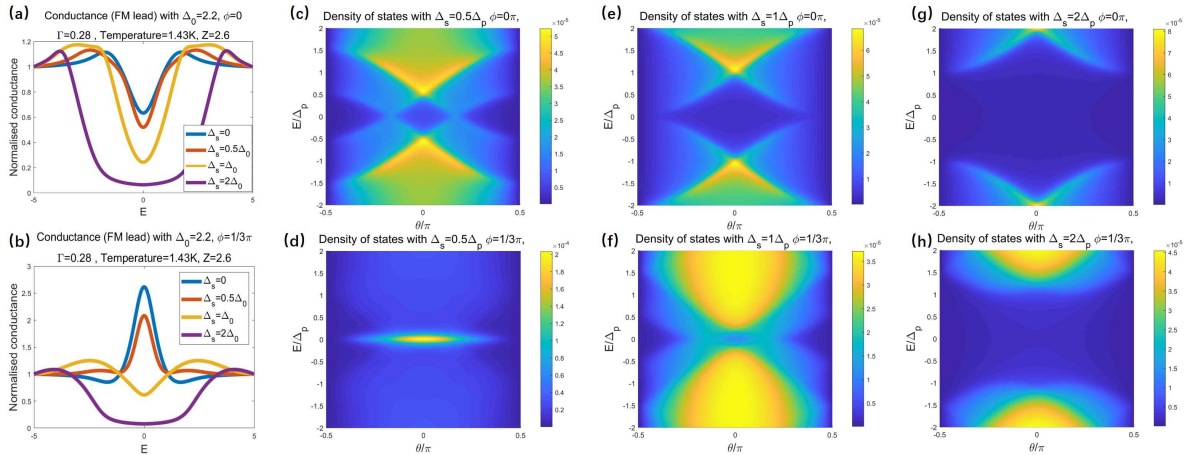}}
 \caption{\raggedright Conductance with different $\Delta_s$ at (a)$\phi=0^0$ and (b)$\phi=60^0$;  density of states for different $\Delta_s$ and (c)$\Delta_s=0.5\Delta_p$, (e)$\Delta_s=\Delta_p$ and (g)$\Delta_s=2\Delta_p$ at $\theta=0^0$ and (d)$\Delta_s=0.5\Delta_p$, (f)$\Delta_s=\Delta_p$ and (h)$\Delta_s=2\Delta_p$ at $\theta=60^0$.
At $\phi$ is 0,increasing the S-wave component widened and deepened the valley in the conductance profile by widening the density-of-states energy gap.
 At $\phi=60^0$ and when the $S$-wave component is lower than the $p$-wave component, the surface state and conductance at zero energy gradually disappear. At $\phi=60^0$ and when the $S$-wave component exceeds  the $p$-wave component, and is further increased, a density of states  energy gap forms near the zero energy, and  the conductance valley widens.
  \label{fig3}}
\end{figure*}
\section{mixture of ABM state and $S$-wave state}\label{conductance-p+s}
The $S$-wave and $p$-wave states can mix in the Bi/Ni bilayer because the inversion symmetry broken\cite{sato2017topological,bauer2012non}.
We next studied the Andreev reflection of unconventional superconductors composed of the ABM state and $S$-wave superconductors.

The Andreev conductance was qualitatively consistent with that of the pure ABM state when $\Delta_s<\Delta_p$ [Fig ~\ref{fig3}(a) and (b)], but with that of the pure $S$-wave state when $\Delta_s\geq\Delta_p$.
As the conductivity at zero energy is mainly contributed by  the zero energy surface state,  we next investigated the effect of the $S$-wave component on the surface state.

As $S$-wave component increased, the surface and bulk states near the zero energy gradually weakened and eventually disappeared [Fig ~\ref{fig3}(c)~(h)].
When $\phi=0^o$, the $S$-wave component  split the funnel-like formation  of the previous density of states into two parts: a left and a right part.
As the $S$-wave component increased, the split widened and the two nodes at zero energy gradually move apart and disappeared, thereby reducing the zero-energy conductance.
The splitting increased the density of states in the overlapping parts of the two nodes.
Increasing  the $S$-wave component, also shifted the higher density region from the zero energy, increasing the conductance valley width.
However, at $\phi=60^o$, as the $S$-wave component increased from 0 to the $p$-wave component, the surface state  weakened and disappeared.
When the $S$-wave component exceeded the $p$-wave component, it split the density of states into two parts.
Further increases of the $S$-wave component gradually increased the distance between the two parts and  diminished the conductance at zero energy, widening the conductance valley.
Moreover, in this case, the $S$-wave component more severely affected the  conductance in FM lead than conductance in NM lead.

In summary, the $S$-wave component reduced the surface state and the conductance at zero energy. When the $S$-wave component was less than the $p$-wave component, the conductance profile resembled that of the pure ABM state and also matched the experimental results\cite{zhao2018triplet}.
The energy band retained it zero-energy nodes  in this case.
However, when the S-wave component exceeded the p-wave component, the conductance profile formed a shape of the valley and the energy band formed a globe gap. Therefore, when the $S$-wave component was small, the conductance was qualitatively consistent with the point contact results\cite{zhao2018triplet} but the energy band failed to explain the time-domain THz spectroscopy\cite{chauhan2019nodeless}.
In contrast, when the S-wave component was  large, the energy band was  consistent with the time-domain THz spectroscopy\cite{chauhan2019nodeless} but the conductance failed to explain the point contact results\cite{zhao2018triplet}.

\section{Summary}
We studied the Andreev reflection conductance between ferromagnetic lead and  two types of superconductors (a pure ABM state superconductor and a mixed state ABM state and $S$-wave state) by the BTK function method.
First, we found that the conductance of the pure ABM state is consistent with that of point contact experiments\cite{zhao2018triplet}.
Second, the result of the mixed state with a small $S$-wave component was qualitatively consistent with that of the pure ABM state and the point contact experiments\cite{zhao2018triplet}.
However,when the $S$-wave component was large, the conductance deviated from the point results\cite{zhao2018triplet} because the gap opened and widened in the energy band.
We also calculated the local density of states, and attributed the conductance peak at zero energy to the surface state of the ABM state component.
Our work provides some complementary explanations for the results of recent experiments.
\section{Acknowledgments}
This work is supported by the National Natural Science
Foundation of China (No.11674028, No.61774017, No.11734004 and No.21421003) and National Key Research and Development Program of China(Grant No. 2017YFA0303300).
\bibliography{andreev}
\section{appendix A: surface green function}
Here, we briefly introduce the method of calculating the surface Green's function\cite{0305-4608-14-5-016}.
First, we discretize the Hamiltonian along the z direction and label each layer with its corresponding  z value (layer i=1~n).
Then, the Hamiltonian is given by:
\begin{equation}
H=\left(\begin{array}{ccccc}
H_{0l,0l} & H_{0l,1l}\\
H_{1l,0l} & H_{0l,0l}\\
 &  & \ddots & \ddots\\
 &  & \ddots & H_{0l,0l} & H_{0l,1l}\\
 &  &  & H_{1l,0l} & H_{0l,0l}
\end{array}\right),
\end{equation}
where $H_{il,i'l}$ denotes the coupling between the i and i' layer.
After discretizing the Hamiltonian, the surface Green's function is obtained by the following procedure:

First define the parameters:
\begin{equation}
\begin{array}{c}
\alpha_{0}=(\omega-H_{0l,0l})^{-1}H_{1l,0l}\\
\beta_{0}=(\omega-H_{0l,0l})^{-1}H_{0l,1l}
\end{array}
\end{equation}
Second, iterate the expressions until $\alpha_{i}\rightarrow0$, $\beta_{i}\rightarrow0$
\begin{equation}
\begin{array}{c}
\alpha_{i}=(1-\alpha_{i-1}\beta_{i-1}-\beta_{i-1}\alpha_{i-1})^{-1}\alpha_{i-1}^{2}\\
\beta_{i}=(1-\alpha_{i-1}\beta_{i-1}-\beta_{i-1}\alpha_{i-1})^{-1}\beta_{i-1}^{2}
\end{array}
\end{equation}
Third. define $T=\alpha_{0}+\beta_{0}\alpha_{1}+\cdots+\beta_{0}\beta_{1}\cdots\beta_{n-1}\alpha_{n}$
The surface Green's function is :$g_{00}=\left\{ \omega-H_{0l,0l}-H_{0l,1l}T\right\} ^{-1}$.
From this function, we obtain the density of states on the the superconductor surface.
\section{appendix B: Fitting of the experimental results}
\begin{figure}
\scalebox{0.4}[0.4]{\includegraphics[0,0][580,250]{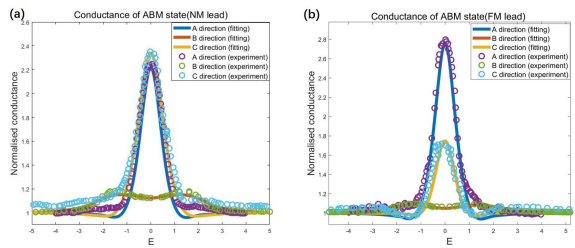}}
 \caption{\raggedright  Fitting of the experimental conductance-energy profiles in different types of lead: FM(right) and NM(left).
 Here,  the order parameter in  the pure ABM state is calculated by  Eq \ref{abm}.
  \label{figa1}}
\end{figure}

Using our model developed in Section~\ref{model} we fited the experimentally obtained normalized Andreev reflection conductances in FM and NM lead  (see Fig ~\ref{figa1} ). The fittings were obtained different incident surfaces with different interface parameters and different $\phi$.

 For the results of NM lead, the  parameters were set as follows: $\Delta_0=2.2,Z=2.2,\Gamma=0.2,T=1.43K,\phi=0.25\pi$ in the a plane, $\Delta_0=2.2,Z=0.64,\Gamma=0.33,T=1.43K,\phi=0.0\pi$ in the b plane, and $\Delta_0=2.2,Z=1.8,\Gamma=0.2,T=1.43K,\phi=0.3\pi$ in the c plane.
For the results of FM lead, the parameters were set to:$\Delta_0=2.0,Z=1.8,\Gamma=0.0.05,T=1.43K,\phi=0.3\pi$ in the a plane, $\Delta_0=2.0,Z=0.698,\Gamma=0.6,T=1.43K,\phi=0.0\pi$ in the b plane, and $\Delta_0=2.0,Z=2.6,\Gamma=0.28,T=1.43K,\phi=0.21\pi$ in the c plane.

\section{appendix C: The conductance with different incidence plane}

\begin{figure}
\scalebox{0.4}[0.4]{\includegraphics[0,0][607,440]{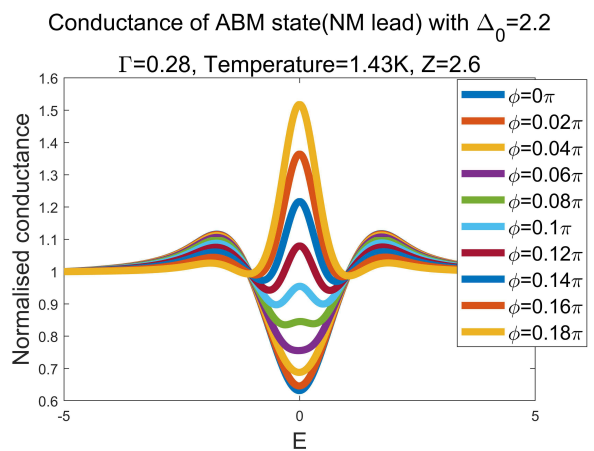}}
 \caption{\raggedright  Conductance versus energy profiles at different incidence planes. The conductance varies slowly  when the incident plane changes.
  \label{figa2}}
\end{figure}

 Using our model, we calculated the normalized the Andreev reflection  conductance for different incidence planes of NM lead.  The results are plotted in Fig ~\ref{figa2}. The conductance varied slowly with changes in the incident plane.
\section{appendix D: influence of $S$-wave component  on energy band, conductance and surface states}
\begin{figure*}
\scalebox{0.9}[0.9]{\includegraphics[41,47][565,547]{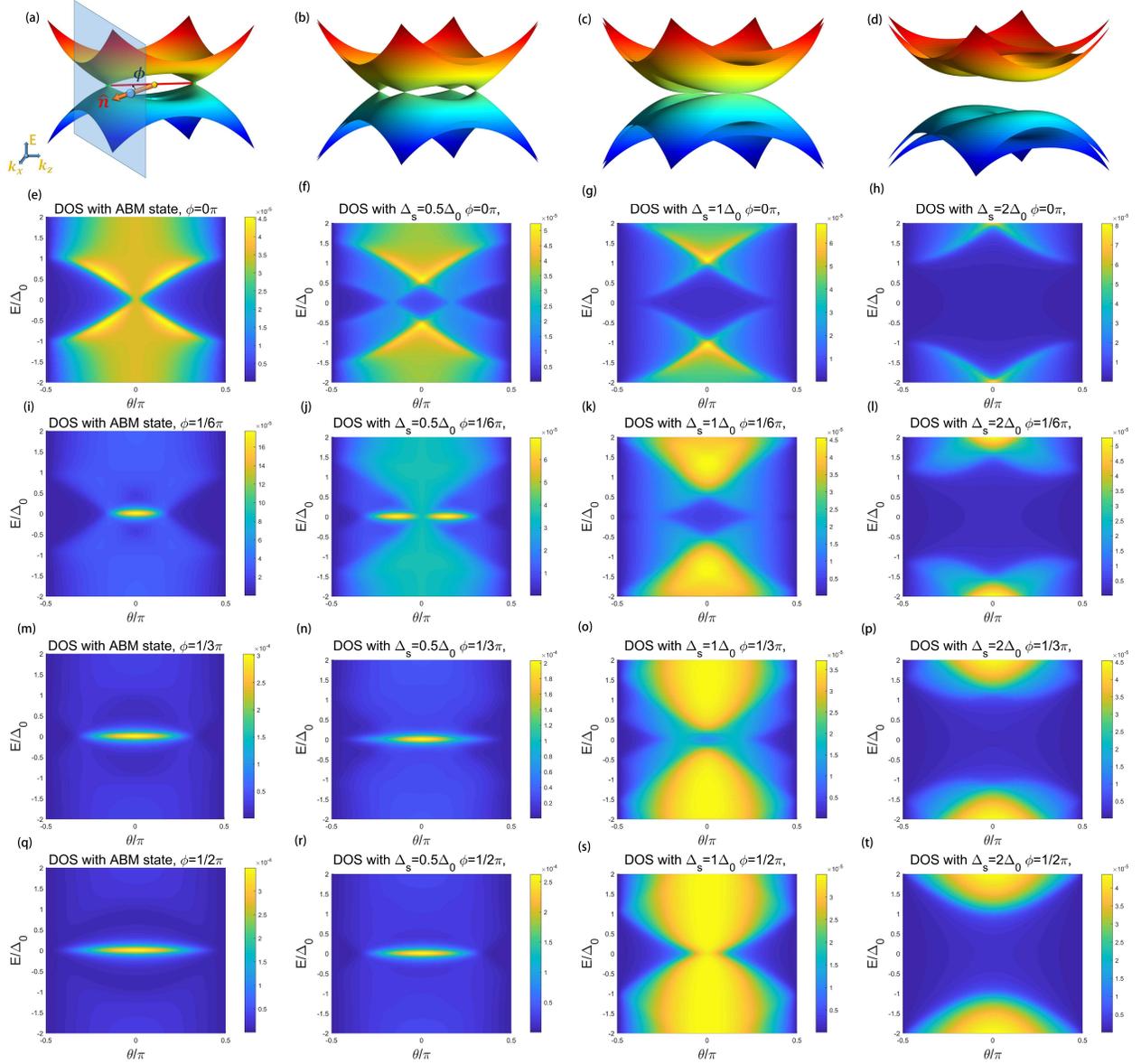}}
 \caption{\raggedright Influence of $\Delta_s$ on the band structure and local density of states for different incident faces between the FM lead and superconductor.(a)-(d) shown the energy band in pure ABM states, $\Delta_s=0.5\Delta_p$, $\Delta_s=\Delta_p$ and $\Delta_s=2\Delta_p$ respectively; (e)-(h),(i)-(l),(m)-(p) and (q)-(t) show the  local density of states when $\phi=0^0$,$\phi=30^0$,$\phi=60^0$,and $\phi=90^0$ in those superconductivity states.
  The $\Delta_s$  decreases the conductance near the zero energy and suppresses the surface state. When $\Delta_s\geqslant\Delta_p$ and $\phi>0$ the surface state and conductance peak vanished.
  \label{figa3}}
\end{figure*}

We first calculated the density of states for different $S$-wave components and incident planes. The S-wave exerted a  huge influence on the energy band and density of states, splitting both original energy band [Fig~\ref{figa3}(a)-(d)] and the original hourglass profile [Fig ~\ref{figa3}(e)-(g)] into two parts.

As shown in Fig ~\ref{figa3}(i)-(k), (m)-(o) and (q)-(s), the $S$-wave component also decreased the surface density of states. At $\Delta_s=\Delta_p$, the surface states disappeared completely.

Moreover, the $S$-wave component created a growing gap in both the energy band and density of states for different incident planes  [Fig~\ref{figa3}(c)and(d), (k)and(l), (o)and(p), (s)and(t)].


\begin{figure*}
\scalebox{0.9}[0.9]{\includegraphics[43,52][540,610]{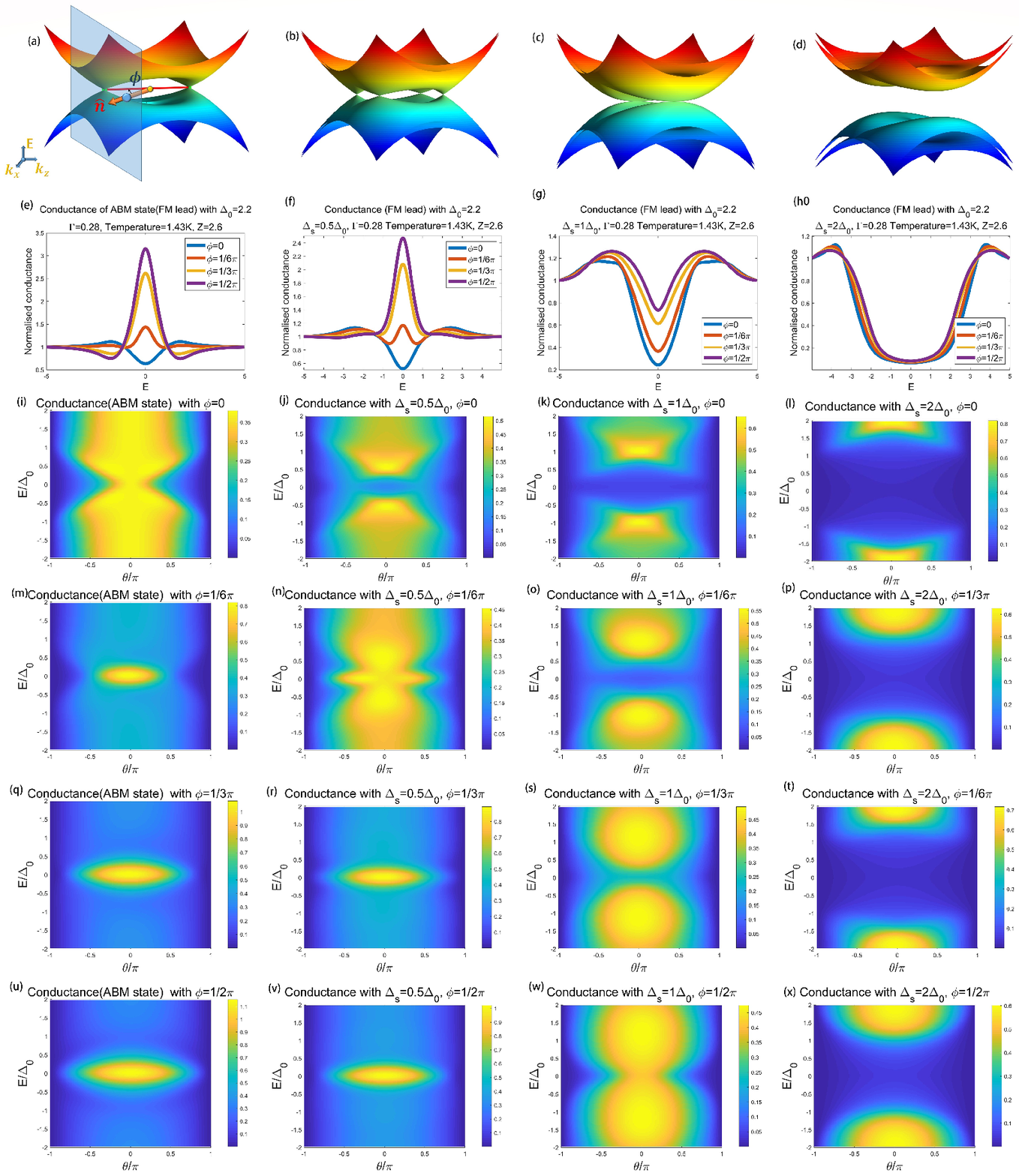}}
 \caption{\raggedright Influence of $\Delta_s$ on the conductance and its spectrum for different incident planes between the FM lead and superconductor.
 (a)-(d)Show the schematic diagram of energy band in pure ABM states, $\Delta_s=0.5\Delta_p$, $\Delta_s=\Delta_p$ and $\Delta_s=2\Delta_p$ respectively; (e)-(h) show the conductance in those superconductivity states; (i)-(l),(m)-(p),(q)-(t) and (u)-(x) show the  conductance spectrum when $\phi=0^0$,$\phi=30^0$,$\phi=60^0$,and $\phi=90^0$ in those superconductivity states.
  The $\Delta_s$ decreases the conductance near the zero energy region. When $\Delta_s\geqslant\Delta_p$ the conductance peak vanishes.
  \label{figa4}}
\end{figure*}

Second, we calculated the influence of the $S$-wave component on the conductance for different incident planes between the FM lead and superconductor.
As shown in  Fig ~\ref{figa4}(e)-(h), the $S$-wave component decreased the conductance near the zero energy.
When $\Delta_s<\Delta_p$, the conductance profile resembled that of  Andreev conductance in the pure ABM state; when $\Delta_s\geqslant\Delta_p$, it was similar to the  Andreev conductance in a pure $S$-wave superconductor.

The influence of the $S$-wave component on the conductance spectrum for different incident planes is shown in  Fig ~\ref{figa4}(e)-(x). The  conductance spectrum resembled that of the local density of states. When $\Delta_s<\Delta_p$ and $\phi$ was non-zero, the $S$-wave component decreased the conductance near the zero energy by decreasing the surface density of states near the zero energy.
However, when $\Delta_s<\Delta_p$ and $\phi=0$, it decreased the conductance near the zero energy by splitting  the density of states.
Finally, when $\Delta_s\geqslant\Delta_p$, it decreased the conductance far from the zero energy by expanding the gap between the high density of states regions.

Note that the conductance spectra and densities of states are consistent at very small energy expansions ($\Gamma$).

Next, we calculated the influence of the $S$-wave component on the conductance for different incident planes between NM lead and superconductor.
The conductance and its spectrum were different in magnitude but qualitatively consistent with those of FM lead  in the same situations.

\end{document}